\documentclass[preprint,aps,amsmath,superscriptaddress,nofootinbib]{revtex4}

\usepackage{dcolumn}
\usepackage{bm}
\usepackage{epsfig}
\usepackage{graphicx}
\usepackage{amsmath}
\usepackage{amssymb}
\usepackage{mathrsfs}
\usepackage{color}
\usepackage[usenames,dvipsnames]{xcolor}
\usepackage{hyperref}
\usepackage{braket}
\usepackage{multirow}
\usepackage{array}
\usepackage{slashed}
\usepackage{rotating}
\usepackage{feynmp-auto}
\usepackage{tikz}
\usepackage{caption}
\usepackage{subcaption}

\usepackage{amsfonts}

\def\be{\begin{equation}}
\def\ee{\end{equation}}
\def\bea{\begin{eqnarray}}
\def\eea{\end{eqnarray}}

\newcommand{\pv}{{\bf p}}

\newcommand{\qv}{{\bf q}}

\newcommand{\kv}{{\bf k}}
\newcommand{\bv}{{\bf b}}

\newcommand{\comment}[1]{}

\def\slashchar#1{\setbox0=\hbox{$#1$}           
   \dimen0=\wd0                                 
   \setbox1=\hbox{/} \dimen1=\wd1               
   \ifdim\dimen0>\dimen1                        
      \rlap{\hbox to \dimen0{\hfil/\hfil}}      
      #1                                        
   \else                                        
      \rlap{\hbox to \dimen1{\hfil$#1$\hfil}}   
      /                                         
   \fi}

\begin{document}

\title{Transverse Momentum Dependent PDFs in Chiral Effective Theory}


\author{Marston Copeland}
\email{paul.copeland@duke.edu}
\affiliation{Department of Physics, Duke University, Durham, North Carolina\ 27708, USA\\}

\author{Thomas Mehen}
\email{mehen@duke.edu}
\affiliation{Department of Physics, Duke University, Durham, North Carolina\ 27708, USA\\}

\begin{abstract} 
We develop a theoretical framework to match transverse momentum dependent parton distribution functions (TMD PDFs) onto chiral effective theory operators. In this framework the TMD PDF is expressed as a convolution of TMD hadronic distribution functions, which describe fluctuations of initial states into intermediate hadrons in chiral perturbation theory, and short distance matching coefficients, which are the TMD PDFs of intermediate hadrons in the chiral limit. The various limits of the matching condition are explored and an operator product expansion is applied to the high energy TMD matching coefficients, allowing them to be written in terms of the collinear valence PDFs of intermediate hadrons. As an example, we calculate the isovector TMD hadronic distribution functions for the proton at leading order in the chiral expansion.
\end{abstract}

\maketitle

\section{Introduction}

Since the advent of quantum chromodynamics (QCD) there has been a considerable effort by nuclear and particle physicists to precisely determine the parton distribution functions (PDFs) and fragmentation functions of hadrons. These quantities appear in factorization theorems of QCD \cite{Collins:1989gx, Collins:2011zzd, Bauer:2002nz, Collins:1981va} and encode the nonperturbative behavior of quarks and gluons in hadrons. Their nonperturbative nature means that traditional perturbative QCD can not be used to calculate them from first principles. 

In recent decades a lot of effort has centered around studying transverse momentum dependent PDFs (TMD PDFs or TMDs, for short) \cite{Ji:2004wu, Collins:1984kg, Meng:1995yn, Mulders:1995dh, Accardi:2012qut, AbdulKhalek:2021gbh}. The TMD PDFs are generalizations of collinear PDFs that contain unique information about a parton's momentum in directions transverse to the collision axis. They provide a richer picture of the distribution of partons within the nucleon and include correlations between parton transverse momentum, parton polarization, and hadron polarization that are not accessible in collinear factorization \cite{Mulders:1995dh, Boussarie:2023izj, Bacchetta:2006tn}. TMDs have been studied in great detail by many authors, see Ref. \cite{Boussarie:2023izj} for a review. Like their collinear siblings, these quantities are also nonperturbative. 

Recently, effective field theories (EFTs) have been successfully applied in numerous instances to extract information about TMD distributions. EFTs exploit the symmetries of QCD and develop expansions in small ratios of scales, allowing for systematic calculations in the appropriate energy regimes. Recent applications of EFT to TMD distributions include the calculation of TMD fragmentation functions for heavy hadrons in heavy quark effective theory \cite{vonKuk:2023jfd, vonKuk:2024uxe, Dai:2023rvd} and for heavy quarkonium in non-relativistic QCD \cite{Echevarria:2020qjk, Echevarria:2023dme, Copeland:2023wbu, Copeland:2023qed}. These applications exploit the fact that the heavy quark mass provides a perturbative scale, making some aspects of the corresponding TMDs calculable. Another recent application is the use of Soft Collinear Effective Theory \cite{Bauer:2000ew, Bauer:2000yr, Bauer:2001ct, Bauer:2001yt} to study subleading TMDs \cite{Ebert:2021jhy} (for a QCD analysis of subleading TMDs, see Ref. \cite{Gamberg:2022lju}). 

For light hadrons like pions and nucleons, the relevant effective theory for describing low energy interactions is chiral perturbation theory ($\chi$PT). $\chi$PT is an EFT that exploits the chiral symmetry breaking pattern of QCD to describe the long distance behavior of light hadrons. Many hadronic properties have successfully been studied using $\chi$PT. Some recent examples include hadron masses \cite{Scherer:2002tk, Young:2009zb, Copeland:2020ljp, Copeland:2021qni, Owa:2023tbk}, scalar charges \cite{Copeland:2021qni, Young:2009zb, Owa:2023tbk}, and form factors \cite{Shanahan:2014cga, Shanahan:2014tja, CSSM:2014knt, Wang:2022bxo, Alharazin:2023zzc}. The many applications of $\chi$PT are too great to list, however for a broad overview see Refs. \cite{Scherer:2002tk, Scherer:2009bt}.

Additionally, while PDFs are fundamentally short distance objects, they can receive long distance contributions from, for example, a nucleon fluctuating into a pion and a baryon, meaning that $\chi$PT can be used to study these effects as well. In fact, $\chi$PT has been extensively applied to collinear PDFs \cite{Chen:2001nb, Arndt:2001ye, Burkardt:2012hk, Shanahan:2013xw, Ji:2013bca, Salamu:2014pka, Salamu:2018cny, Salamu:2019dok, Wang:2022bxo} to study sea quark distributions in the nucleon and explain the $ \bar{d}-\bar{u}$ asymmetry observed in the proton \cite{NewMuon:1993oys, HERMES:1998uvc, NuSea:2001idv}. $\chi$PT has even been used to study light cone distribution functions \cite{Chen:2003fp} and generalized parton distributions \cite{Ando:2006sk, Moiseeva:2012zi, He:2022leb, Strikman:2009bd}. 

TMD PDFs, however, have not been carefully considered in $\chi$PT, despite their current popularity in the literature \cite{Boussarie:2023izj}. In fact, to our knowledge, only one of the eight TMD PDFs in the proton (the Sivers function) has been studied so far in a chiral effective framework. In Ref. \cite{He:2019fzn} the authors use a nonlocal chiral effective theory with a finite range regulator, offering some theoretical insight to how TMDs are described in chiral effective theory. In this paper, we match TMD PDFs onto operators in conventional $\chi$PT, which we expect to provide a description of the low-energy physics of TMD PDFs. 



When matching the collinear PDF onto $\chi$PT, it is very useful that the moments of the collinear PDF are matrix elements of local operators in QCD. Matching local operators in QCD onto local operators in $\chi$PT is a straightforward procedure. The matching coefficients of the chiral operators are moments of the PDFs computed at short distance in the chiral limit. It is easy to construct a non-local operator whose moments correspond to the aforementioned local chiral operators, and it follows that the full PDF is a convolution of the short distance PDF with the matrix elements of the nonlocal chiral operators, which we call hadronic distribution functions. When generalizing this to the TMD PDF, we assume that the convolution formula holds with both the PDF and the chiral operator having separation in transverse position. We give arguments to support this hypothesis. This enables us to calculate chiral corrections to the TMD PDFs, as has been done for the collinear PDFs in the past. 

In Section \ref{sec: Def} we review chiral effective theory and how it has been applied to collinear PDFs in previous work. Building off of these ideas, we introduce the TMD hadronic distribution functions and their convolution with the valence TMD PDFs of the intermediate hadron states. In Sec. \ref{sec: TMD split func} we calculate the nucleon TMD hadronic distribution functions at next-to-leading order (NLO) and discuss the properties of the operators. Then, in Sec. \ref{sec: results} we study these hadronic distribution functions numerically as functions of the transverse momentum. Finally, in Sec. \ref{sec: conclusion}, we conclude and summarize next steps. There is also an appendix that discusses a sum rule that is satisfied by the integrated TMD (and collinear) PDFs which serves as an important check of our calculations.
\section{Definitions and framework}
\label{sec: Def}
\subsection{SU(2) chiral effective theory}
To begin we review relativistic SU(2) chiral effective theory. The Lagrangian for the theory is \cite{Scherer:2002tk},
\begin{equation}
    {\cal L} = {\bar N}( i {\slashed D} -M ) N + g_A{\bar N} \gamma^\mu \gamma_5 u_\mu N + \frac{f_\pi^2}{4}{\rm Tr}\big[\partial^\mu\Sigma^\dagger \partial_\mu \Sigma].
\end{equation}
Here, the nucleon fields are in a doublet, $ N = (\begin{smallmatrix} p\\ n \end{smallmatrix}$), and the sigma fields are,
\begin{equation}
\Sigma = \exp\bigg(i\frac{\sqrt{2}M}{f_\pi}\bigg), ~~ M = \frac{\tau^a \pi^a}{\sqrt{2}} = 
\begin{pmatrix}
\pi^0/\sqrt{2}  & \pi^+ \\
\pi^- & -\pi^0/\sqrt{2}
\end{pmatrix}.
\end{equation}
The chirally covariant derivative and axial vector field are defined as 

\begin{equation}
\begin{aligned}
     D_\mu = \partial_\mu + \Gamma_\mu, ~~~ &~~~~\Gamma_\mu = \frac12(\xi^\dagger\partial_\mu \xi + \xi \partial_\mu \xi^\dagger)\\
     u_\mu = &\frac{i}{2}(\xi^\dagger\partial_\mu \xi - \xi \partial_\mu \xi^\dagger),
\end{aligned}
\label{eq: chiral defs}
\end{equation}
where $\xi^2 = \Sigma$.
The fields transform like 
\begin{equation}
\begin{aligned}
    N(x) \to U(x) N(x), ~~~ & \xi(x) \to L \xi(x) U^\dagger(x) = U(x) \xi(x) R^\dagger,\\
    &\Sigma(x) \to L\Sigma R^\dagger, 
\label{eq: Chiral transformations}
\end{aligned}
\end{equation}
under $SU(2)_L \times SU(2)_R$ transformations. Note the transformation $U(x)$ is local while the $L$ and $R$ transformations are global and do not depend on the position of the fields. The transformation properties of the fields will be important when constructing the nonlocal operators used for the TMD hadronic distribution functions. 

We note that heavy-baryon chiral perturbation theory (HB$\chi$PT) is an alternative version of the theory we are using. The advantage of HB$\chi$PT is that the large baryon mass is removed from the theory, so that all derivatives are proportional to the low-energy scales and power counting is manifest. This is the effective theory used in Refs. \cite{Chen:2001nb,Arndt:2001ye} when studying collinear PDFs, for example. However, it has been argued that relativistic chiral effective theory offers somewhat better convergence in low-energy perturbation theory \cite{Pascalutsa:2005nd, Bernard:1995dp, Becher:1999he}. Moreover, it also has a well defined power-counting scheme, as long as the low-energy constants are appropriately renormalized \cite{Pascalutsa:2005nd, Gegelia:1999qt}. Since the matrix elements of PDFs and TMD PDFs are customarily expressed in terms of relativistic lightcone operators, matching onto the relativistic version of $\chi$PT is more straightforward. For these reasons, we elect to use relativistic $\chi$PT.

\subsection{Parton Distribution Functions}

Even though parton distributions are well established in the literature, we review their definitions and list our conventions in this section for convenience. In this paper we use light front coordinates, defining two light-like vectors $n$ and $\bar{n}$ such that for a vector $v$, $n\cdot v = v^+ = \frac{1}{\sqrt{2}}(v^0 +v^3)$ and $\bar{n}\cdot v = v^- = \frac{1}{\sqrt{2}}(v^0 - v^3)$. The transverse components are denoted as ${\bf v}_T = (v_x, v_y)$.

The parton distribution function for an unpolarized parton $i$ in a hadron $X$ is defined as,
\begin{equation}
    f_{i/X}(\zeta) = \int \frac{db^-}{2 \pi} e^{-i \zeta b^- P^+}\bra{X} \overline{\psi}_i(b^-) W_n(b^-;0,-\infty) \frac{\slashed{n}}{2} W_n^{\dagger} (0; 0,-\infty) \psi_i(0)\ket{X},
\end{equation}
where $\zeta$ is the partonic momentum fraction of the hadron's light-like momentum, $P^+$, and $b^-$ is the light-like separation of the quark fields. For the sake of clarity, we omit the dependence on the renormalization scale. In practice, we are matching the bare PDFs onto the low-energy effective theory operators since the hard perturbative corrections to the PDF occur at scales above what $\chi$PT can describe. The Wilson lines are defined by the path-ordered exponential,
\begin{equation}
    W_n (x; a, b) =  {\cal P} \, {\rm exp}\left[-i g_0 \int_a^b ds~ n\cdot A^{a} (x + s n) t^a\right].
\end{equation}
and are included to maintain gauge invariance in the operator. The distributions we work with in this paper will not be flavor singlets. The nonsinglet distribution is,
\begin{equation}
    f^a_{i/X}(\zeta) = \int \frac{db^-}{2 \pi} e^{-i \zeta b^- P^+}\bra{X} \overline{\psi}(b^-) W_n(b^-;0,-\infty) \Gamma^a\frac{\slashed{n}}{2} W_n^{\dagger} (0; 0,-\infty) \psi(0)\ket{X},
\end{equation}
where the field $\psi$ is an isodoublet and $\Gamma^a$ is a matrix in  isospin space. Likewise, the nonsinglet TMDs in position space are defined as \cite{Boussarie:2023izj}
\begin{equation}
    {\tilde f}^a_{i/X}(\zeta, \bv_T) =  \int \frac{db^-}{2 \pi} e^{-i \zeta b^- P^+}\bra{X} \overline{\psi}(b)W_\sqsubset(b, 0) \Gamma^a \frac{\slashed{n}}{2} \psi(0)\ket{X},
\label{eq: nonsinglet TMD}
\end{equation}
where $b = (b^-, 0, {\bf b}_T)$ so there is now a transverse separation, ${\bf b}_T$. 
As in the collinear case, we omit the dependence on the renormalization scale and also the Collins-Soper scale. We use the tilde to denote that the TMD PDF is in position space. The staple shaped Wilson line is given by 
\begin{equation}
\begin{aligned}
    W_\sqsubset(b,0) = &W_n ( b; 0,-\infty) W_{\hat{b}_T}(-\infty n ; b_T, 0) W^\dagger_n ( 0 ;0,-\infty)
\end{aligned}
\end{equation}
and is again included to maintain gauge invariance in the operator. In this paper, we only consider TMD PDFs with unpolarized parton in an unpolarized nucleon. However, it would be straightforward to generalize this work by matching onto the analogs of Eq. (\ref{eq: nonsinglet TMD}) with spin dependent Dirac structure. $\slashed{n}\gamma_5/2$ and $\sigma^{+ \alpha}\gamma_5/2$ \cite{Boussarie:2023izj}. 
\subsection{Collinear convolution formalism}

In effective field theory, QCD operators are matched on low-energy, effective theory operators that obey the same symmetries and transformation properties. This procedure is well defined for local operators where the matching coefficients simply multiply the effective operators. But what about for nonlocal operators? How should the coefficients and effective operators depend on the separation of fields? This is especially confusing for TMDs, where the quark fields are not just separated in one collinear direction, $b^-$, but in the transverse direction as well. To gain some insight into how one might match TMDs onto chiral effective theory, we review the collinear convolution formalism introduced by Chen and Ji \cite{Chen:2001nb}.



Their matching procedure exploits a property of the PDF which allows it to be expanded in terms of {\it local} twist-two operators. Expanding in the $b^-$ variable, we can write the collinear PDF as
\begin{equation}
\begin{aligned}
    f^a_{q/X}(\zeta) = \int \frac{db^-}{2 \pi} e^{i \zeta b^- P^+}\sum_k \frac{(-i b^-)^{k-1}}{(k-1)!}  \braket{({\cal O^+})^{k, a}},
\end{aligned}
\label{eq: PDF OPE}
\end{equation}
where
\begin{equation}
    \braket{({\cal O^+})^{k,a}} = \bra{X} \overline{\psi}\frac{\slashed{n}}{2} \Gamma^a (i n\cdot D)^{k-1} \psi\ket{X}.
\end{equation}
These local twist two operators can be matched onto chiral effective operators with the same quantum numbers \cite{Chen:2001nb, Arndt:2001ye},  
\begin{equation}
    ({\cal O^+})^{k,a} = \sum_H c^k_{qH} {\cal O}^{k,a}_H
\label{eq: twist2 matching}
\end{equation}
where the matching coefficients, $c^k_{qH}$, describes the high energy physics at the scale $p^2 \ge \Lambda_\chi^2$ and the hadronic operators ${\cal O}^k_H$ in the low energy effective theory describe physics at the scale $p^2 \ll \Lambda_\chi^2$ \cite{Chen:2003fp}. 

\comment{
Since chiral operators transform under left and right handed transformations and are not necessarily in flavor singlets, it is more convenient to work with the projections,
\begin{equation}
\begin{aligned}
    f^{L, a}_{i/X}(\xi) = & \int \frac{db^-}{2 \pi} e^{i b^- (\xi P^+)}\bra{X} \overline{\psi}(b^-) W_n(b^-)\Gamma^a \slashed{n} \frac{(1-\gamma_5)}{2}W^{\dagger} (0) \psi(0)\ket{X}\\
    f^{R, a}_{i/X}(\xi) = & \int \frac{db^-}{2 \pi} e^{i b^- (\xi P^+)}\bra{X} \overline{\psi}(b^-) W_n(b^-)\Gamma^a \slashed{n} \frac{(1+\gamma_5)}{2}W^{\dagger} (0) \psi(0)\ket{X}.
\end{aligned}
\end{equation}
The original PDFs can easily be recovered using $f^{a}_{i/X}(\xi) = f^{R, a}_{i/X}(\xi) + f^{L, a}_{i/X}(\xi)$. Likewise, one could access the longitudinally polarized quark PDFs via $g^{a}_{i/X}(\xi) = f^{R, a}_{i/X}(\xi) - f^{L, a}_{i/X}(\xi)$.}
In SU(2) chiral effective theory, we want to match onto operators that transform in the same way as the PDF under $SU(2)_L \times SU(2)_R$ transformations. The leading order operators with the same quantum numbers and symmetries are,
\begin{equation}
\begin{aligned}
    & {\cal O}^{k,a}_{\pi} =\frac{f_\pi^2}{ 8}{\rm Tr}\big[\Sigma^\dagger \Gamma^a (i n\cdot \partial)^k \Sigma + \Sigma \Gamma^a (i n\cdot \partial)^k \Sigma^{\dagger}], \\
    & {\cal O}^{k,a}_{N}  = \frac14 \bar{N} \slashed{n} \big[\xi^\dagger \Gamma^a(i n\cdot \partial)^{k-1}  \xi + \xi\Gamma^a (i n\cdot \partial)^{k-1}  \xi^{\dagger}\big] N,\\
    & {\cal O}^{A; k,a}_{N}= \frac{g_A}{4}\bar{N} \slashed{n}\gamma_5 \big[\xi^\dagger\Gamma^a (i n\cdot \partial)^{k-1}  \xi - \xi \Gamma^a (i n\cdot \partial)^{k-1} \xi^{\dagger}\big] N,
\label{eq: ChiPT moments}
\end{aligned}
\end{equation}
where the fields are all located at the same space-time point. The first and second operator normalizations are fixed to give a factor of ($P^+$)$^k$ at leading order in the chiral expansion when their matrix elements are evaluated between single pion and nucleon states, respectively, where $P^+$ is the momentum of the external hadron. The third operator has no tree level matrix element between single particle states so there is no analogous way to fix its normalization. Instead, we fix the last operator's normalization by comparing with the operator 
\begin{equation}
\begin{aligned}
\Delta {\cal O}^{A; k,a}_{N}= \frac{g_A}{4}\bar{N} \slashed{n}\gamma_5 \big[\xi^\dagger\Gamma^a (i n\cdot \partial)^{k-1}  \xi + \xi \Gamma^a (i n\cdot \partial)^{k-1} \xi^{\dagger}\big] N,
\label{eq: ChiPT moments}
\end{aligned}
\end{equation}
(which matches onto the longitudinally polarized PDF, $g_1$, at leading order) since parity and $SU(2)_L \times SU(2)_R$ symmetry constrain the coefficients of ${\cal O}^{A; k,a}_{N} $ and $\Delta {\cal O}^{A; k,a}_{N}$ to be the same \cite{Moiseeva:2012zi}. In both of the normalizations ${\cal O}^{A; k,a}_{N} $ and $\Delta {\cal O}^{A; k,a}_{N}$, we also include a factor of $g_A$ for convenience. To compensate, the matching coefficients of these operators should be divided by $g_A$. The normalizations in Eq. (\ref{eq: ChiPT moments}) are also necessary for the operators to satisfy a nontrivial sum rule, which is discussed more in Appendix \ref{app: sum rule}. Note that the hadronic operators are color singlet objects and are thus SU(3) gauge invariant. More importantly, they are also chirally invariant, as can be seen by applying Eq. \ref{eq: Chiral transformations}. 

The hadronic operators between an external state $X$ define hadronic distribution functions in terms of their moments \cite{Chen:2001nb},
\begin{equation}
    \braket{{\cal O}^{k,a}_H} = (P^+)^k\int d\beta \beta^{k-1} f^a_{HX}(\beta).
\label{eq: splitting moment}
\end{equation}
%
Inverting the Mellin transform gives an explicit form for the nonlocal hadronic distribution functions,
\begin{equation}
    f^a_{HX}(\beta) = \int \frac{db^-}{2\pi} e^{-i b^-\beta P^+} \bra{X} {\cal O}^a_H (b^-, 0) \ket{X} ,
\end{equation}
where the nonlocal hadronic operators are given by 
\begin{equation}
\begin{aligned}
    & {\cal O}^a_{\pi}(b^-, 0) =\frac{f_\pi^2}{ 8}{\rm Tr}\big[\Sigma^\dagger(b^-) \Gamma^a (i n\cdot \partial) \Sigma(0) + \Sigma(b^-) \Gamma^a (i n\cdot \partial) \Sigma^{\dagger}(0)], \\
    & {\cal O}^a_{N}(b^-,0)  = \frac14 \bar{N}(b^-) \slashed{n} \big[\xi^\dagger(b^-) \Gamma^a  \xi(0) + \xi(b^-)\Gamma^a \xi^{\dagger}(0)\big] N(0),\\
    & {\cal O}^{A; a}_{N}(b^-, 0)  = \frac{g_A}{4}\bar{N}(b^-) \slashed{n}\gamma_5 \big[\xi^\dagger(b^-) \Gamma^a  \xi(0) - \xi(b^-) \Gamma^a \xi^{\dagger}(0)\big] N(0).
\label{eq: ChiPT operators}
\end{aligned}
\end{equation}
These hadronic distribution functions are also called light-cone momentum distributions \cite{Chen:2001nb, Burkardt:2012hk, Ji:2013bca} of intermediate hadrons in a final state and collinear hadronic splitting functions \cite{Salamu:2018cny, Salamu:2019dok} the literature.
Interestingly, the operators in Eq. (\ref{eq: ChiPT operators}) are still chirally invariant because the $\xi(x)$ fields transform to cancel out the $U(x)$ dependence from the nucleon's transformation. 

%

A different way to arrive at Eq. (\ref{eq: ChiPT operators}) is to pick nonlocal operators with the same quantum numbers as the PDF. This was observed when constructing nonlocal operators to match onto nucleon GPDs \cite{Moiseeva:2012zi}. To see how this works, remember that the transformation properties are given by Eq. (\ref{eq: Chiral transformations}), so the combinations
\begin{equation}
\begin{aligned}
    [\xi^\dagger(x) N(x)] \to R [\xi^\dagger(x) N(x)] , ~~~ [\xi(x) N(x)] \to L [\xi(x) N(x)] ,
\label{eq: N building blocks}
\end{aligned}
\end{equation}
transform independently of the spatial coordinate, $x$.

Now, using Eq. (\ref{eq: N building blocks}), chirally invariant combinations of nonlocal nucleon fields can be constructed
\begin{equation}
\begin{aligned}
     \bar{N}(x)(\xi^\dagger(x) \Gamma \xi(y) \pm \xi(x) \Gamma \xi^\dagger(y)] N(y)
\end{aligned}
\end{equation}
where $\Gamma$ is some structure consisting of Dirac gamma matrices, isospin matrices, and/or derivatives to match the quantum numbers of the original operator. Likewise, the combinations
\begin{equation}
    {\rm Tr}\big[\Sigma^\dagger(x) (i n\cdot \partial) \Sigma(y) \pm \Sigma(x) (i n\cdot \partial) \Sigma^{\dagger}(y)\big]
\end{equation}
are also chirally invariant. In both cases, the $\pm$ is chosen to match the parity of the PDF operator. 

It is worth pointing out that the PDF operator, $\bar{\psi}(b^-) W_n(b^-,0)\slashed{n}\psi(0)$, is a nonlocal generalization of the vector current in QCD. In effective field theories, it is standard practice to match Noether currents from one theory onto another. The isospin-1 vector Noether currents of the chiral Lagrangian are of the form of Eq. (\ref{eq: ChiPT operators}) with $b^- = 0$. If we try to turn that current into a nonlocal operator with nucleons or pion fields living at different spacetime points, there is a unique way to do that if one wishes to maintain chiral invariance. In this sense, for the PDF we are matching the nonlocal QCD vector current onto its analog in $\chi$PT.  


As a side note, we remark that Eq. (\ref{eq: N building blocks}) is reminiscent of Wilson lines in QCD, which are included to preserve gauge invariance for nonlocal operators with quark fields, except here the necessary symmetry is not gauge invariance, but instead chiral invariance. A QCD combination invariant under gauge transformations is 
\begin{equation}
\begin{aligned}
     W_n^{\dagger} (x; 0, -\infty) \psi(x)
\label{eq: psi building blocks}
\end{aligned}
\end{equation}
since the field transforms like $\psi(x)\to U_n(x) \psi(x)$ and the Wilson line transforms like $W_n^{\dagger} (x;0, -\infty) \to W_n^{\dagger}  (x;0, -\infty) U_n^\dagger(x)$ (after imposing the boundary condition $U_n(-\infty) = 1)$. This can be compared with the chiral combinations in Eq. (\ref{eq: N building blocks}). 

\comment{
The $\xi$ fields also have somewhat similar properties to the Wilson lines - for example, in the same manner that the Wilson lines can define the gauge covariant derivative,
\begin{equation}
    W_n(x) (in\cdot \partial) W^\dagger_n(x) {\cal O}= i n\cdot D_n {\cal O},
\end{equation}
the $\xi$ fields relate to the chirally covariant derivative by,
\begin{equation}
\begin{aligned}
    \frac12[\xi(x) \partial_\mu \xi^\dagger (x){\cal O} + \xi^\dagger(x)\partial_\mu \xi (x) {\cal O}]= \big[\partial_\mu + \Gamma_\mu\big] {\cal O}
\end{aligned}
\end{equation}
where $\Gamma_\mu$ is defined in Eq. (\ref{eq: chiral defs}).
}

For the high energy part of the matching in Eq. (\ref{eq: twist2 matching}) we can determine a physical interpretation  for the matching coefficients as well. When the intermediate ($H$) and external ($X$) hadronic states are the same, the moments of hadronic distribution functions produce delta functions and powers of the hadrons momentum at leading order. Hence, the matching coefficients can be identified with the PDF moments of the hadron $H$ in the chiral limit \cite{Chen:2001nb, Chen:2003fp, Moiseeva:2012zi},
\begin{equation}
    c^k_{qH} = \int^1_{-1} d\alpha \alpha^{k-1} q^{(0)}_{H}(\alpha).
\label{eq: matching coeff}
\end{equation}
The $q_H^{(0)}$'s should be normalized accordingly to compensate for the normalizations of the operators in Eq. (\ref{eq: ChiPT moments}).

Putting Eqs. (\ref{eq: splitting moment}) and (\ref{eq: matching coeff}) into Eq. (\ref{eq: PDF OPE}), we can produce the convolution, 

\begin{equation}
\begin{aligned}
    f^a_{q/X}(\zeta) =& \sum_H \bigg[\int_{|\zeta|}^1 \frac{d \alpha}{\alpha} q^{(0)}_{H} (\alpha) f^a_{HX} \bigg(\frac{\zeta}{\alpha}\bigg)-\int^{-|\zeta|}_{-1} \frac{d \alpha}{\alpha} q^{(0)}_{H} (\alpha) f^a_{HX} \bigg(\frac{\zeta}{\alpha}\bigg)\bigg]\\
    =& \sum_H \int_{\zeta}^1 \frac{d \alpha}{\alpha} q^{v}_{H} (\alpha) f^a_{HX} \bigg(\frac{\zeta}{\alpha}\bigg)
\label{eq: col. match}
\end{aligned}
\end{equation}
where $q^{v}_{H}(\alpha) = q^{(0)}_{H}(\alpha) + q^{(0)}_{H}(-\alpha)$ is referred to as the valence PDF in the intermediate hadron \cite{Wang:2022bxo}. 


\subsection{TMD matching}

The derivation of Eq. (\ref{eq: col. match}) is possible because the matrix elements of twist-2 operators are equal to a power of the collinear momentum of the external states times the moments of the PDF. There is no analogous relation for the moments of the TMD PDFs in terms of local operators, making it difficult to use this approach. Recent advances in the study of TMD moments \cite{delRio:2024vvq} may be helpful for solving this problem, but we save this for future studies.
Instead, we use a different property of the nonlocal operators to conjecture the TMD matching condition. In Eq. (\ref{eq: N building blocks}) we observed that certain combinations of the chiral fields and nucleons transform globally under left or right transformations.The chiral theory is only concerned about the flavor transformations which happen at the endpoints of the nonlocal PDF operator where the quark fields are located. The Wilson lines do not undergo any such transformations and are thus not described by the chiral operators. Thus we can match left and right handed quark fields dressed by Wilson lines directly onto the chiral building blocks located at the same point, 
\begin{equation}
\begin{aligned}
    & W^\dagger_{\hat{b}_T}(-\infty n ; 0, -\infty) W^\dagger_n ( 0 ; 0, -\infty) \psi_L(0) \to c_q(0) {\cal O}_L^\chi(0)\\
    & W^\dagger_{\hat{b}_T}(-\infty n ; 0, -\infty) W^\dagger_n ( 0 ; 0, -\infty) \psi_R(0) \to c_q(0) {\cal O}_R^\chi(0)
\end{aligned}
\end{equation}
with the same argument applying for $\bar{\psi}(b)W_n ( b; 0, -\infty) W_{\hat{b}_T}(-\infty n ; b_T, -\infty)$. The matching coefficients depend on the position of the fields and $c^\dagger_q(b)c_q(0)$ can be identified with the TMD distributions in the intermediate hadron. From this observation, we can argue that the TMD matching is given by,  

\begin{equation}
    \tilde{f}^a_{q/X}(\zeta, \bv_T) = \sum_H \int_\zeta^1 \frac{d \alpha}{\alpha} \tilde{q}^{v}_{H} (\alpha, \bv_T) \tilde{f}^a_{HX} \bigg(\frac{\zeta}{\alpha}, \bv_T\bigg),
\label{eq: TMD matching}
\end{equation}
%
%
where we now have a convolution between the valence TMD PDF in the intermediate hadron, $\tilde{q}_{H}^{v}(\alpha, \bv_T)$, and a {\it TMD hadronic distribution function},
\begin{equation}
    \tilde{f}^a_{HX}(\beta, \bv_T) = \int \frac{db^-}{2\pi} e^{-i b^-\beta P^+} \bra{X} {\cal O}^a_H (b, 0) \ket{X} .
\label{eq: TMD split func}
\end{equation}
Here, the TMD chiral operators are given by the generalization of Eq. (\ref{eq: ChiPT operators}) with light-like and transverse separation. For $b = (b^-, 0, {\bf b}_T)$, we have
\begin{equation}
\begin{aligned}
    & {\cal O}^a_{\pi}(b, 0) = \frac{f_\pi^2}{ 8}{\rm Tr}\big[\Sigma^\dagger(b) \Gamma^a (n\cdot \partial) \Sigma(0) + \Sigma(b) \Gamma^a (n\cdot \partial) \Sigma^{\dagger}(0)], \\
    & {\cal O}^a_{N}(b, 0)  = \frac14 \bar{N}(b)\slashed{n} \big[\xi^\dagger(b)\Gamma^a  \xi(0) + \xi(b)\Gamma^a \xi^{\dagger}(0)\big] N(0),\\
    & {\cal O}^{A;a}_{N}(b, 0)  = \frac{g_A}{4}\bar{N}(b) \slashed{n}\gamma_5 \big[\xi^\dagger(b) \Gamma^a  \xi(0) - \xi(b) \Gamma^a \xi^{\dagger}(0)\big] N(0).
\end{aligned}
\label{eq: TMD chiral operators}
\end{equation}
Again, these operators are invariant under $SU(2)_L \times SU(2)_R$ transformations because of the transformation properties of the $\xi(x)$ fields. Notice, the valence TMDs and the TMD hadronic distribution functions both depend on the transverse separation, $\bv_T$.

\comment{ In Eqs. (\ref{eq: TMD chiral operators}) it appears that we have lost information unique to TMDs, such as the Sivers effect, because the chiral operators only contain information about fields containing flavor, i.e., the quarks and not the staple shaped Wilson lines. We note, however, that the effect of the staple shaped Wilson line is to remove time reversal invariance as a symmetry of the TMD PDF. Thus, we can reproduce the Sivers effect by matching onto additional chiral operators that are time reversal odd. Using only nucleons, pions, Dirac structures, and the transverse separation, $\bv_T$, we can write one additional T-odd operator,
\begin{equation}
    {\cal O}^S_{N}(b,0) = i M \bar{N}(b)\epsilon_T^{\mu \nu} \bv_{T ,\mu} \gamma_\nu \gamma_5 \big[\xi^\dagger(b)\Gamma^a  \xi(0) + \xi(b)\Gamma^a \xi^{\dagger}(0)\big] N(0).
\end{equation}
As before, we are only matching the TMD PDF with an unpolarized parton onto the effective theory. If one allowed for polarized partons then additional operators would be needed. }

One important feature of Eq. (\ref{eq: TMD matching}) is that it enforces momentum conservation when a Fourier transform is performed,
\begin{equation}
    f_{q/X}^a(\zeta, \kv_T) = \sum_H \int d^2\pv_T  d^2\qv_T  \int_\zeta^1 \frac{d \alpha}{\alpha} q^{v}_{H} (\alpha, \pv_T) f^a_{HX} \bigg(\frac{\zeta}{\alpha}, \qv_T\bigg)\delta^{(2)}(\pv_T + \qv_T - \kv_T).
\label{eq: TMD matching momentum}
\end{equation}
Physically, this states that the total transverse momentum, $\kv_T$, is the sum of the transverse momentum of the intermediate hadron with respect to the initial state, $\qv_T$, and the transverse momentum of the parton inside of the intermediate hadron, $\pv_T$.

In the convolution of Eq. (\ref{eq: TMD matching momentum}) we are not making any assumption  ion the size of $\kv_T$ because both the effective theory TMD hadronic distribution functions and the matching coefficients depend on the transverse momentum. This is the most general thing to write down for moderate $\kv_T$, as it allows for transverse momentum contributions from both the quarks inside of the intermediate hadron and from the fluctuations into intermediate hadron states.

However, in a regime where the total transverse momentum is small ($\kv_T \sim m_\pi \sim \Lambda_{QCD}$) we would expect the majority of the transverse momentum to come from the low energy hadronic fluctuations (e.g., a proton splitting into a pion and neutron). This is because the valence TMD distributions are matching coefficients, so they encode the high energy physics of the system, i.e., physics at scales $\ge \Lambda_\chi \sim$ 1-2 GeV, which are much larger than the hadronic scales of $\Lambda_{QCD} \sim 200$ MeV. In this small $\kv_T$ scenario, we can imagine that the transverse momentum dependence is described entirely by the effective theory operators, i.e., the TMD hadronic distribution functions, in which case Eq. (\ref{eq: TMD matching momentum}) reduces to, 
\begin{equation}
    f_{q/X}^a(\zeta, \kv_T) = \sum_H \int_\zeta^1 \frac{d \alpha}{\alpha} q^{v}_{H} (\alpha) f^a_{HX} \bigg(\frac{\zeta}{\alpha}, \kv_T\bigg).
\label{eq: Low energy matching}
\end{equation}
This equation could have been produced from the start by assuming that at low $\kv_T$ or large $\bv_T$, the high energy matching coefficients have nothing to say about the transverse momentum dependence. A somewhat similar result was found by Ref. \cite{vonKuk:2023jfd} in heavy quark effective theory when studying heavy quark TMD fragmentation functions in the limit $\kv_T \sim \Lambda_{QCD} \ll m_Q$, where $m_Q$ is the heavy quark mass.

In the opposite limit, when $\kv_T \gg \Lambda_{QCD} \sim m_\pi$, the $\chi$PT operators should not be able to describe the transverse momentum dependence because it is a large scale. In this case, we can treat the TMD PDF as perturbative and expand in $\Lambda_{QCD}^2/\kv_T^2$ \cite{Collins:2011zzd, Boussarie:2023izj}, which effectively takes $\bv_T$ to 0 in both the coefficient and the TMD hadronic distributions. Perturbative corrections will then generate convolutions with high energy coefficients which are found by matching the valence TMD PDFs onto the valence collinear PDFs,
\begin{equation}
\begin{aligned}
    f^a_{q/X}(\zeta, \kv_T) = &\sum_{H, p}  \int_\zeta^1 \frac{d \alpha}{\alpha}\int^1_\alpha \frac{d\sigma}{\sigma}C_{qp}\bigg(\frac{\alpha}{\sigma}, \kv_T\bigg) p_{H}^{v}(\sigma) f^a_{HX} \bigg(\frac{\zeta}{\alpha}\bigg)+{\cal O}\bigg(\frac{\Lambda_{QCD}^2}{\kv_T^2}\bigg) ,
\label{eq: TMD matching HE lim}
\end{aligned}
\end{equation}
where $C_{p q}$ is the perturbatively calculable matching coefficient and $p$/$q$ are the parton subscripts indicating mixing and $p_{H}^{v}(\sigma) $ is the collinear valence PDF corresponding to the intermediate parton $p$ (the $q$ corresponds to the initial quark in the TMD PDF).

For more moderate $\kv_T$, it may be appropriate to treat the high energy coefficient $q_H^v(\alpha, \pv_T)$ as perturbative and expand as above, but also allow for the hadronic distribution function to have transverse momentum as well. This approximation still requires the transverse momentum in the coefficient to be large but the transverse momentum in the TMD hadronic splitting function must be small for chiral perturbation theory to be valid. Switching back to $\bv_T$ space for clarity, this expansion gives, 

\begin{equation}
\begin{aligned}
    \tilde{f}^a_{q/X}(\zeta, \bv_T) = &\sum_{H, p}  \int_\zeta^1 \frac{d \alpha}{\alpha}\bigg[\int^1_\alpha \frac{d\sigma}{\sigma}\tilde{C}_{q p}\bigg(\frac{\alpha}{\sigma}, \bv_T\bigg) \tilde{p}_{H}^{v}(\sigma) +{\cal O}\bigg(\Lambda_{QCD}^2\bv_T^2\bigg) \bigg]\tilde{f}^a_{HX} \bigg(\frac{\zeta}{\alpha}, \bv_T\bigg)
\label{eq: TMD matching OPE}
\end{aligned}
\end{equation}
At leading order in the perturbative expansion, the coefficient $\tilde{C}_{p q}$ is independent of the transverse separation, $\bv_T$, meaning that all of the transverse momentum dependence would stem from the TMD hadronic distribution function. This would be highly predictive. However, as higher order corrections are included, large logarithms in $\bv_T$ begin to appear which need to be resummed to all orders. Unfortunately, this usually introduces some nonperturbative effects from the Collins-Soper kernel, which may or may not be described in this formalism. In fact, there may be some additional nonperturbative effects lurking in $q^{v}_{H} (\alpha, \pv_T)$ in general, despite its high energy nature. These effects may need to be modelled in some way and convoluted with the TMD hadronic distribution functions to describe the full TMD PDF.

For arbitrary $\bv_T$ or $\kv_T$, implementing the convolution in Eq. (\ref{eq: TMD matching}) will likely not be straightforward phenomenologically. Some combination of the various ansatzs, models, and limits described above will be needed to piece together a full description of data. The variations of this framework will be explored in detail by comparing against state of the art extractions of TMD PDFs in future work. It is worthwhile to point out that all of the TMD convolutions presented in this paper reproduce the collinear convolution at leading order in perturbation theory after integrating over the transverse momentum. This is because the collinear hadronic distribution function is simply the integrated TMD hadronic distribution function. 


\section{TMD hadronic distribution functions}
\label{sec: TMD split func}
%
To actually calculate the TMD hadronic distribution functions, it is necessary to expand the $\xi$ and $\Sigma$ exponentials in terms of the pion fields. As we will see, we find that having the pion fields at different spacetime points will lead to interesting features in the calculations. 

\subsection{Operators}

We begin with the unpolarized nucleon operator. The terms that could contribute up to ${\cal O}(p_\pi^2/f_\pi^2, m_\pi^2/f_\pi^2)$ are
\begin{equation}
\begin{aligned}
     {\cal O}^a_{N}(b, 0)  = &\frac12 \bar{N}(b) \slashed{n} \bigg[\Gamma^a + \frac{1}{4f_\pi^2}\bigg( \tau\cdot\pi(b) \Gamma^a \tau\cdot\pi(0) \bigg)\\
     &- \frac{1}{8f_\pi^2}\bigg((\tau\cdot \pi(b))^2\Gamma^a + \Gamma^a (\tau\cdot \pi(0))^2\bigg) + {\cal O}\bigg(\frac{p_\pi^3}{\Lambda_\chi^3}, \frac{m_\pi^3}{\Lambda_\chi^3}\bigg)\bigg]N(0),\\
\end{aligned}
\label{eq: ON}
\end{equation}
Notice there are three types of terms. For clarity, we refer to the term with pion fields located at two different points as the symmetric tadpole operator, and the terms with pion fields at the same point as asymmetric tadpoles.

Next, for the axial operator, the leading order contribution is
\begin{equation}
\begin{aligned}
     {\cal O}^{A; a}_{N}(b, 0)  = &\frac{-ig_A}{4f_\pi}\bar{N}(b) \slashed{n} \gamma_5 \bigg[\bigg( \tau\cdot\pi(b) \Gamma^a -  \Gamma^a \tau\cdot\pi(0)\bigg) + {\cal O}\bigg(\frac{p_\pi^3}{\Lambda_\chi^3}, \frac{m_\pi^3}{\Lambda_\chi^3}\bigg)\bigg]N(0).
\end{aligned}
\label{eq: OA}
\end{equation}
Here, we seemingly have two terms, one pion field located at $b$, and another pion field located at $0$.

Finally, we have the purely pionic operator, ${\cal O}_{\pi}(b,0)$. The first terms that contribute from the operator are
\begin{equation}
\begin{aligned}
     {\cal O}^{a}_{\pi}(x, y) =  \frac14 &{\rm Tr}\bigg[\bigg( \tau\cdot\pi(b) \Gamma^a (i n\cdot \partial) \tau\cdot\pi(0)\bigg)- \frac12\Gamma^a (i n\cdot \partial)(\tau\cdot \pi(0))^2 + {\cal O}\bigg(\frac{p_\pi^3}{\Lambda_\chi^3}, \frac{m_\pi^3}{\Lambda_\chi^3}\bigg)\bigg]
\label{eq: Opi}
\end{aligned}
\end{equation}
The last term is a total derivative and hence vanishes because the initial and final state momenta in a PDF are the same. There are more terms up to the order we are working (i.e., up to 1/$f_\pi^2$). However, these terms have three or four pion fields, so they vanish between single nucleon states without additional interactions. 

Note that because these currents are nonlocal and the pion fields are often at different space time points, the isospin structures do not simplify neatly like they would if all fields were at the same point. Chiral corrections to these operators can  be computed by using the lowest order interaction Lagrangian in chiral effective theory. This procedure yields analytic expressions for the hadronic distribution functions, which can be implemented in Eq. (\ref{eq: TMD matching OPE}). We give an example in the following sections. 
%

\subsection{TMD hadronic distribution functions for the nucleon}
\begin{figure}[t]
\includegraphics[width = .7\linewidth]{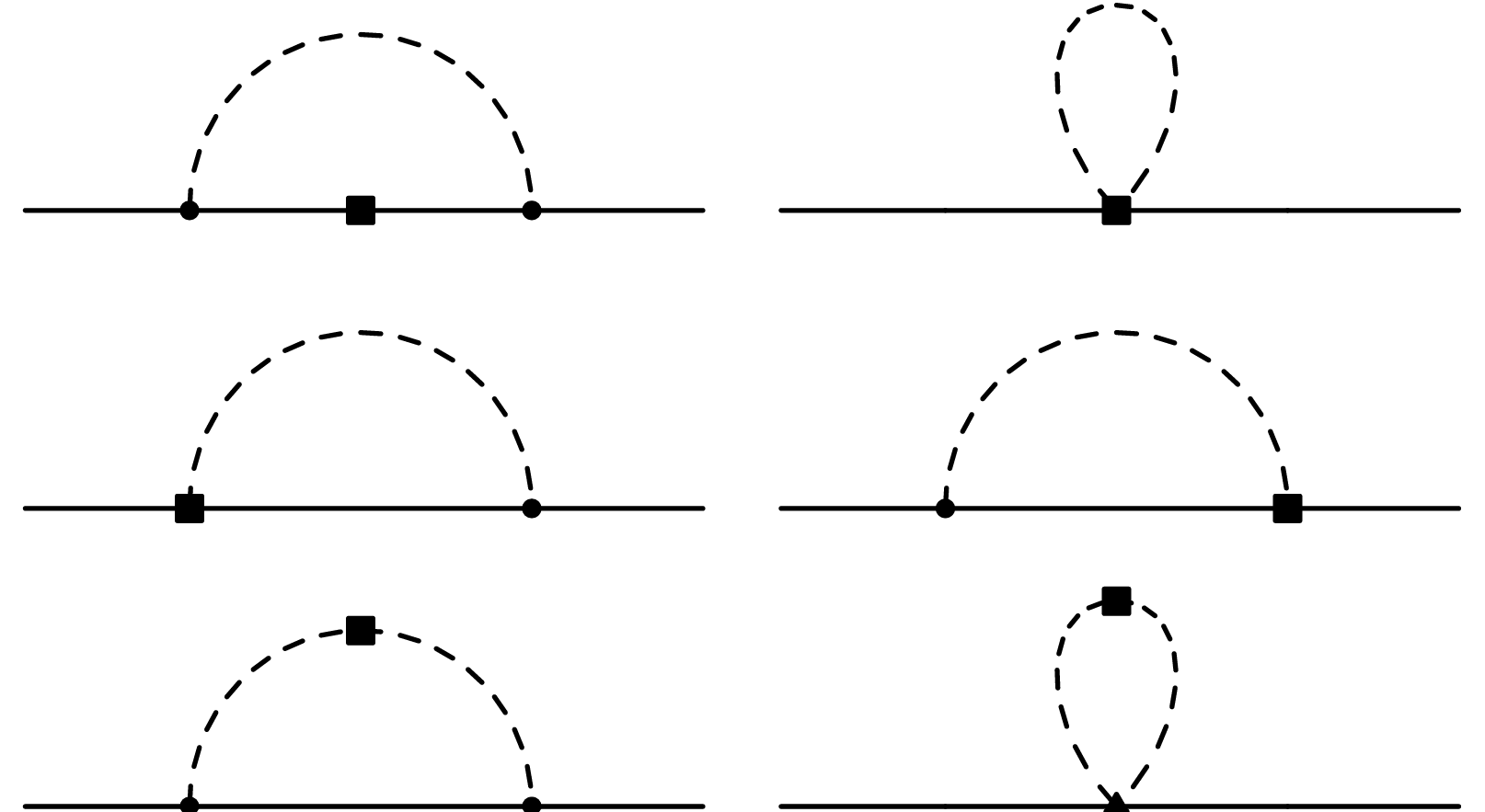}

    \caption{Loop corrections to all operators. Squares represent an operator insertion, circles are an insertion of the axial interaction, and the triangle is the
    $\pi \pi N N$ interaction. }
    \label{fig: operator diagrams}
\end{figure}
In this work, we only consider nucleon external states, so for our purposes the relevant interaction Lagrangian is,
\begin{equation}
    {\cal L}_{int} = \frac{i}{2}{\bar N}\gamma^\mu(\xi^\dagger\partial_\mu \xi + \xi \partial_\mu \xi^\dagger) N +i\frac{g_A}{2}{\bar N} \gamma^\mu \gamma_5 (\xi^\dagger\partial_\mu \xi - \xi \partial_\mu \xi^\dagger) N .
\end{equation}
which when expanded gives
\begin{equation}
    {\cal L}_{int} = -\frac{1}{4f_\pi^2}{\bar N}\gamma^\mu \epsilon^{a b c} \tau^c (\pi^a \partial_\mu \pi^b) \ N -\frac{g_A}{2f_\pi}{\bar N} \gamma^\mu \gamma_5 \partial_\mu (\tau\cdot\pi) N .
\end{equation}
%
Chiral corrections to the operators generate the diagrams shown in Fig. \ref{fig: operator diagrams}. Since the operators are nonlocal, most of these diagrams can be represented more accurately by separating the operators (the squares) into two points, 0 and $b$, then attaching pion lines to different locations. These variations are presented below with the detailed discussion of the calculations.

The calculations have various isospin factors which depend on the operator we are considering and the choice of the $\Gamma^a$ matrix. We calculate the hadronic distribution functions in the most general manner possible, and denote isospin factors with $I_{HX}^a$ (where the $HX$ and $a$ are indicated by the hadronic distribution function $f_{HX}^a$).

Since we are working with nonlocal operators, we will also need to conserve momentum at the $b$ operator. A Feynman rule for this can be derived as follows. Generically, we have (Fourier transformed) operators of the form,

\begin{equation}
    \int \frac{db^- d^2 \bv_T}{(2\pi)^3} e^{-i\beta P^+ b^-} e^{i\bv_T \cdot \qv_T} \bra{N} {\cal O}(b) {\cal O}(0) \ket{N}.
\end{equation}
This is equivalent to 
\begin{equation}
    \int \frac{db^- d^2 \bv_T}{(2\pi)^3} e^{-i\beta P^+ b^-} e^{i\bv_T \cdot \qv_T} \bra{N} e^{ib\cdot{\hat {\cal P}}}{\cal O}(0) e^{-ib\cdot{\hat {\cal P}}}{\cal O}(0) \ket{N}
\end{equation}
by translational invariance, where ${\hat {\cal P}}$ is the momentum projection operator. After acting the $e^{ib\cdot{\hat {\cal P}}}$ on the bra, $\bra{N}$, we can rewrite our operators in the practical form of,
\begin{equation}
    \bra{N} {\cal O}(0) \delta^{(2)}({\hat  {\cal P}}_T - \qv_T)\delta(n\cdot{\hat {\cal P}} - (1-\beta )P^+){\cal O}(0) \ket{N}.
\label{eq: delta function rule}
\end{equation}
Here, the ${\hat {\cal P}}$'s project out any additional, intrinsic, momentum created by higher order chiral perturbation theory corrections. In other words, it projects out the momentum flowing through the operator at the point $b$, plus the momentum of the final state. For example, if a pion carries some intrinsic momentum, $k$, and there is $P-k$ momentum flowing out of the point $b$, then the projector would select $-(P-k)$ + $P$ = $k$. So the delta functions for the Feynman rule would be $\delta^{(2)}(\kv_T - \qv_T)\delta(k^+ - (1-\beta )P^+)$.

Before proceeding, we point out that technically the nucleon wave function renormalization diagrams should be included in Fig. \ref{fig: operator diagrams} as well. Except for their application in Appendix \ref{app: sum rule}, we do not study these contributions in this work, however, because they are not novel and do not affect the TMD dependence of the operators. We now use the expanded operators and Feynman rules to calculate the hadronic distribution functions directly. 

\subsection{Nucleon to Nucleon hadronic distribution functions}
\begin{figure}[t]
\begin{subfigure}{0.3\linewidth}
\includegraphics[width = .76\linewidth]{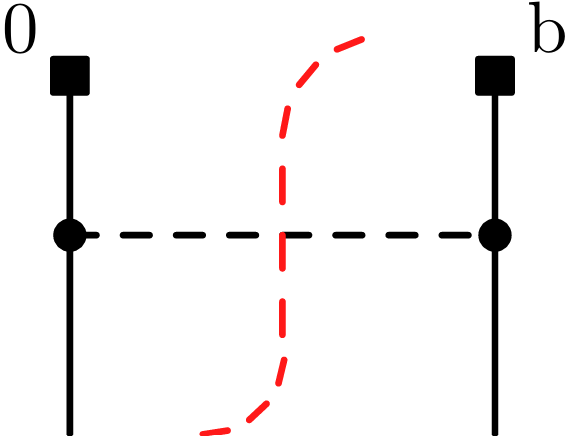}
\caption{}
\label{fig: N LO}
\end{subfigure}
\begin{subfigure}{0.3\linewidth}
\includegraphics[width = .65\linewidth]{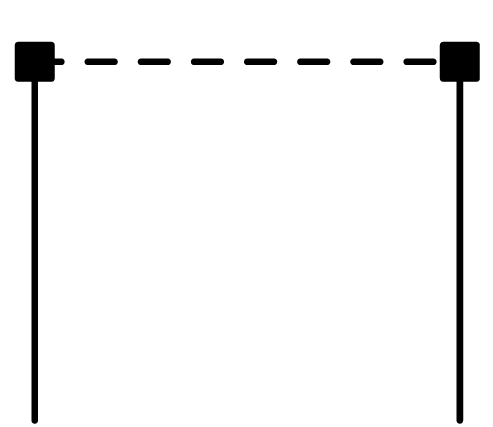}
\caption{}
\label{fig: N sym} 
\end{subfigure}
\begin{subfigure}{0.3\linewidth}
\includegraphics[width = .67\linewidth]{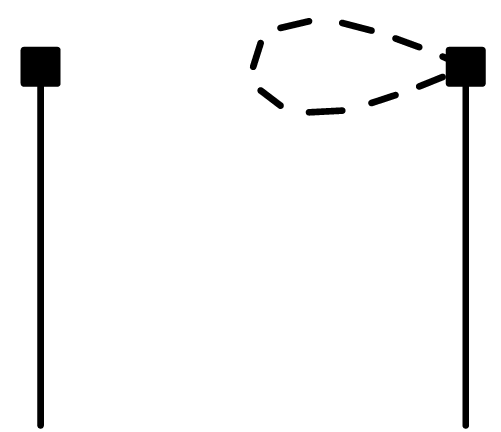}
\caption{}
\label{fig: N off}
\end{subfigure}
    \caption{Loop corrections to nucleon operators. The first diagram represents $f^a_{NN}$, the second is $f^{a, {\rm spl}}_N$, and the last is $f^{a, {\rm tad}}_N$. The mirror of the last diagram is not shown. Squares represent an operator insertion, circles are an insertion of the axial interaction, and the dashed red line represents a cut.}
    \label{fig: NO diagrams}
\end{figure}

%
First, we study the hadronic distribution function defined by nucleon operator, ${\cal O}_N$. At next-to-leading order, we find there are three contributions. These can be calculated directly by inserting the expanded operator, Eq. (\ref{eq: ON}), directly into the definition of the TMD hadronic distribution function in Eq. (\ref{eq: TMD split func}). For convenience, we calculate all quantities in momentum space. The first term in the nucleon operator, $N(b)\slashed{n}\Gamma^a N(0)$, generates Fig. \ref{fig: N LO} by two insertions of the axial interaction. Applying the traditional chiral effective theory Feynman rules and the delta function rule from Eq. (\ref{eq: delta function rule}) gives
\begin{equation}
\begin{aligned}
    f^a_{NN}(\beta, \qv_T) = &\frac{g_A^2 I_{NN}^a}{8 f_\pi^2}\int \frac{d^4 k}{(2\pi)^4} \overline{u}(P) \slashed{k}\gamma_5 \frac{i(\slashed{P}-\slashed{k}+M)}{(P-k)^2 - M^2 +i\epsilon}\frac{i}{k^2-m_\pi^2+i\epsilon} \slashed{n}\\
    &\times \frac{i(\slashed{P}-\slashed{k}+M)}{(P-k)^2 - M^2 +i\epsilon} \gamma_5 \slashed{k} u(P)\delta^{(2)}(\kv_T - \qv_T)\delta(k^+ - (1-\beta )P^+),
\end{aligned}
\label{eq: NN feyn}
\end{equation}
where we have taken the initial nucleon to have only collinear momentum, $P^+$, and no transverse momentum. As explained above, $I_{NN}^a$ is an isospin factor that depends on the choice of $\Gamma^a$ in the operator definition. Equation (\ref{eq: NN feyn}) is what you might expect - it is the collinear nucleon-nucleon hadronic distribution function \cite{Burkardt:2012hk, Wang:2022bxo}, except with a transverse momentum delta function from the transverse separation of the fields. 

To simplify the numerator, we first average over spins and take a trace. After the spin trace, we are left with three integrals,
\begin{equation}
\begin{aligned}
    f^a_{NN}(\beta, \qv_T) = &\frac{-i g_A^2 I_{NN}^a}{16 f_\pi^2}\int \frac{d^4 k}{(2\pi)^4}\bigg[\frac{16M^2 m_\pi^2(P^+-k^+)}{(k^2-m_\pi^2+i\epsilon)[(P-k)^2-M^2+i\epsilon]^2} - \frac{4P^+}{k^2-m_\pi^2+i\epsilon}\\
    &+ \frac{16M^2 (P^+-k^+)}{[(P-k)^2-M^2+i\epsilon]^2}\bigg]\delta^{(2)}(\kv_T - \qv_T)\delta(k^+ - (1-\beta )P^+).
\label{eq: Traced fNN}
\end{aligned}
\end{equation}

Since $k^+$ and ${\bf k}_T$ are constrained by the delta functions, we only really need to integrate over $k^-$. These integrals are given by,

\begin{subequations}
\begin{equation}
    \int dk^-  \frac{1}{(k^2-m_\pi^2+i\epsilon)[(P-k)^2 - M^2+i\epsilon]^2} = \frac{-\pi ik^+}{(P^+)^2[{\bf k}_T^2 +m_\pi^2(1-k^+/P^+) +M^2(k^+/P^+)^2]^2}
\label{eq: DN Dpi2 int}
\end{equation}
\begin{equation}
    \int dk^-  \frac{1}{k^2 - m_\pi^2+i\epsilon} =  i\pi\bigg[-\frac{1}{\epsilon} + \log\bigg(\frac{\kv_T^2 + m_\pi^2}{\mu_\chi^2}\bigg)\bigg] \delta(k^+),
    \label{eq: Dpi km} 
\end{equation}
\begin{equation}
    \int dk^-  \frac{1}{[(P-k)^2 - M^2+i\epsilon]^2} = \frac{i\pi }{\kv_T^2 + M^2}\delta(P^+-k^+),
\label{eq: DN km int}
\end{equation}
\label{eq: km ints}
\end{subequations}
where $\mu_\chi$ is a mass scale \cite{Burkardt:2012hk}. Note, for the first integral we have propagators with poles in two different halves of the complex plane as long as $k^+/P^+$ is between 0 and 1. This allows us to use Cauchy's integral formula to take the pole at $k^- = (k_T^2+m_\pi^2-i\epsilon)/(2k^+)$, which is equivalent to setting the pion on-shell. We interpret this as taking a cut through the diagram and this is depicted in Fig. \ref{fig: N LO} by the dashed red line. This is similar to the one loop calculation of the quark TMD PDF, where the gluon propagator is set on-shell \cite{Boussarie:2023izj}.

To evaluate the integrals in Eqs. (\ref{eq: Dpi km}) and (\ref{eq: DN km int})  use the following trick. First, we observe that if $k^+$ (or for Eq. (\ref{eq: DN km int}), $P^+ -k^+$) is positive or negative, the contour can be deformed around the pole to make the integral vanish. However, as already mentioned, if $k^+$ ($P^+ -k^+$) is 0, then the integral is divergent - meaning the result is proportional to a delta function in $k^+$ ($P^+ -k^+$). To find the proportionality coefficient for Eq. (\ref{eq: Dpi km}), we evaluate 
\begin{equation}
    \int dk^+ dk^- \frac{1}{k^2-m_\pi^2+i\epsilon} 
\end{equation}
using dimensional regularization in $2-2\epsilon$ dimensions. Working with the assumption that the $k^-$ integral by itself must give the $d^2k$ result multiplied with $\delta(k^+)$ lets us produce Eq. (\ref{eq: Dpi km}) and (\ref{eq: DN km int}).
Notice since the last term in Eq. (\ref{eq: Traced fNN}) is proportional to $P^+-k^+$, it vanishes due to the $\delta(P^+-k^+)$ from the integral. 

The integrals in Eqs. (\ref{eq: Dpi km}) and (\ref{eq: DN km int}) are actually ill defined and not readily regulated by traditional methods like dimensional regularization or a cutoff. We are only able to get a result using dimensional regularization when comparing with the $d^{2-2\epsilon}k$ result. It would be interesting to explore whether other methods, like rapidity regulators, are able to systematically control these divergences. A more detailed investigation of this divergence is warranted in the future. 

We note that the divergence in Eq. (\ref{eq: Dpi km}) occurs at order $m_\pi^0$ in the chiral expansion. In other words, it is at the same order as our original operators. Therefore we can remove this divergence by introducing a counterterm proportional to $\overline{N}(b^-) \slashed{n} [\xi^\dagger(b^-) \Gamma^a \xi(0) + H.C.] N(0)$, since this gives a $\delta(k^+) = \delta(1-\beta)$ at tree level.  In general, we use dimensional regularization in the $\overline{MS}$ and remove all $1/\epsilon$ poles using operator renormalization. Since we are working with a relativistic theory, a more sophisticated renormalization scheme, like the extended on-mass-shell (EOMS) renormalization, could be used \cite{Moiseeva:2012zi, Fuchs:2003qc}.



Integrating over the delta functions in $k^+$ and $\kv_T$ then plugging in the integrals from Eq. (\ref{eq: km ints}) into Eq. (\ref{eq: Traced fNN}) gives, 
\begin{equation}
\begin{aligned}
    f^a_{NN}(\beta, \qv_T) = &\frac{ -g_A^2 I_{NN}^a}{16\pi^3 f_\pi^2}\bigg(\frac{M^2m_\pi^2\beta(1-\beta)}{[\qv_T^2 + M^2(1-\beta)^2+ m_\pi^2\beta]^2}+\frac14\log\bigg(\frac{\qv_T^2+m_\pi^2}{\mu_\chi^2}\bigg)\delta(1-\beta)\bigg)
\end{aligned}
\label{eq: final fNN}
\end{equation}
%


%

Next, we study the ${\cal O}(\pi^2)$ terms in the expanded operator, Eq. (\ref{eq: ON}). The symmetric tadpole operator is shown in Fig. \ref{fig: N sym} and contributes,
\begin{equation}
    f^{a, {\rm s}}_{NN}(\beta, \qv_T) =  \frac{I_{NN}^{a, {\rm s}}}{8f_\pi^2}\int \frac{d^4 k}{(2\pi)^4} \overline{u}(P) \slashed{n} \frac{i}{k^2 - m_\pi^2} u(P) \delta(k^+ - (1-\beta)P^+) \delta^{(2)}(\kv_T-\qv_T).
\end{equation}
The integral over the pion propagator is in Eq. (\ref{eq: Dpi km}) and is nonvanishing only as $k^+$ goes to 0. Evaluating the spin trace and using Eq. (\ref{eq: Dpi km}) gives,
\begin{equation}
    f^{a, {\rm s}}_{NN}(\beta, \qv_T) =  \frac{-I_{NN}^{a,{\rm s}}}{64\pi^3 f_\pi^2} \log\bigg(\frac{\qv_T^2 + m_\pi^2}{\mu_\chi^2}\bigg)  \delta(1-\beta).
\end{equation}
Notice that while this operator has $\qv_T$ dependence, it only contributes at the endpoint when $\beta = 1$. 

Lastly, the asymmetric tadpole operators do not have $\qv_T$ dependence and only contribute for $\beta \to 1$. This is Fig. \ref{fig: N off} and for completeness, we write down its contribution. The asymmetric tadpole is given by,
\begin{equation}
    f^{a, {\rm as}}_{NN}(\beta, \qv_T) =  \frac{-I_{NN}^{a,{\rm as}}}{16f_\pi^2}\int \frac{d^4 k}{(2\pi)^4} \overline{u}(P) \slashed{n} \frac{i}{k^2 - m_\pi^2} u(P) \delta((1-\beta)P^+)\delta^{(2)}(\qv_T).
\end{equation}
Since this contribution has no delta function constraints on the $k^+$ or ${\bf k}_T$ components, we need a regularization prescription. Evaluating the $k$ integrals using dimensional regularization gives, 
\begin{equation}
    f^{a, {\rm as}}_{NN}(\beta, \qv_T) =  \frac{m_\pi^2I_{NN}^{a,{\rm as}}}{8(4\pi)^2 f_\pi^2} \bigg[1 - \log\bigg(\frac{m_\pi^2}{\mu_\chi^2}\bigg)\bigg]  \delta(1-\beta)\delta^{(2)}(\qv_T).
\end{equation}
where $\gamma_E \approx 0.577$ is the Euler-Mascheroni constant. The mirror diagram gives the same contribution. 

\subsection{Axial hadronic distribution functions}

\begin{figure}[t]
\begin{subfigure}{0.35\linewidth}
\includegraphics[width = .65\linewidth]{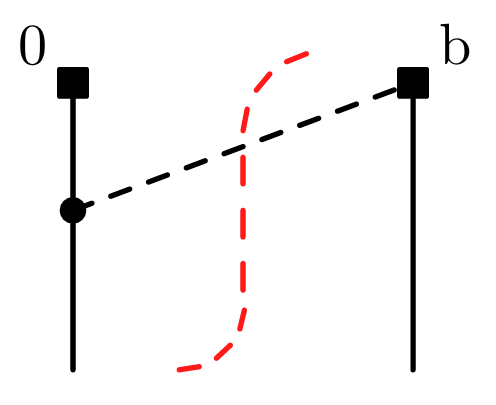}
\caption{}
\label{fig: A cross}
\end{subfigure}
\begin{subfigure}{0.35\linewidth}
\includegraphics[width = .57\linewidth]{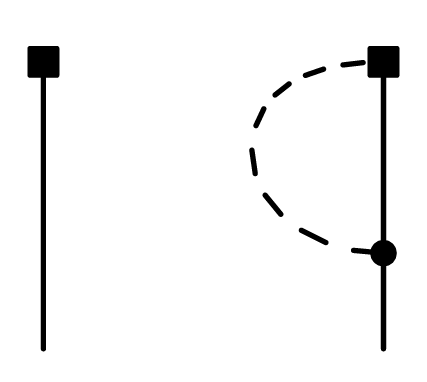}
\caption{}
\label{fig: A loop}
\end{subfigure}
    \caption{Loop corrections to axial operator with cuts. Mirrors are not shown. Squares represent an operator insertion and circles are an insertion of the axial interaction.}
    \label{fig: OA diagrams}
\end{figure}

We saw in Eq. (\ref{eq: OA}) that there are two terms that contribute to the axial operator. One where the pion field is located at $b$ and one where the pion field is at $0$. Each term generates two possible diagrams - the pion either crosses a cut, or reconnects with the nucleon on the same side. Figure \ref{fig: OA diagrams} shows the diagrams with the pions located at $b$. The diagrams with the pion at $0$ are the mirrors of Fig. \ref{fig: OA diagrams}. 

For Fig. \ref{fig: A cross} plus its mirror, the pion crosses a cut giving the contribution
\begin{equation}
\begin{aligned}
    f^{A; a}_{NN}&(\beta, \qv_T) = \frac{-i g_A^2 I_{N N}^{A; a}}{8f_\pi^2}\int \frac{d^4k}{(2\pi)^4}\overline{u}(P)\bigg[\slashed{n}\gamma_5 \frac{i}{k^2-m_\pi^2+i\epsilon}\frac{i(\slashed{p}-\slashed{k}+M)}{(p-k)^2-M^2+i\epsilon}  \slashed{k} \gamma_5\\ 
    &-  \gamma_5 \slashed{k} \frac{i}{k^2-m_\pi^2+i\epsilon}\frac{i(\slashed{p}-\slashed{k}+M)}{(p-k)^2-M^2+i\epsilon} \slashed{n}\gamma_5\bigg]u(P) \delta(k^+ - (1-\beta)P^+)\delta^{(2)}(\qv_T-\kv_T),
\label{eq: axial TMD hadronic distribution functions}
\end{aligned}
\end{equation}
where again, $I_{NN}^{A, a}$ is the isospin factor. Following the same procedure as above, averaging over spins and taking the trace, $ f^{A, a}_{NN}$ can be simplified to 
\begin{equation}
\begin{aligned}
    f^{A; a}_{NN}(\beta, \qv_T) =& - \frac{i g_A^2I_{N N}^{A; a}}{8 f_\pi^2}\int \frac{d^4k}{(2\pi)^4}\bigg[\frac{8k^+M^2}{(k^2-m_\pi^2+i\epsilon)[(P-k)^2-M^2+i\epsilon]}+\frac{4P^+}{k^2-m_\pi^2+i\epsilon}\bigg]\\
    &\times \delta(k^+ - (1-\beta)P^+)\delta^{(2)}(\qv_T-\kv_T).
\end{aligned}
\end{equation}
Plugging in Eq. (\ref{eq: Dpi km}) and using the integral,
\begin{equation}
\begin{aligned}
    \int dk^-  &\frac{1}{(k^2-m_\pi^2+i\epsilon)[(P-k)^2 - M^2+i\epsilon]} \\
    = &\frac{\pi i}{P^+[{\bf k}_T^2 +m_\pi^2(1-k^+/P^+) +M^2(k^+/P^+)^2]},
\label{eq: DpiDN int}
\end{aligned}
\end{equation}
gives the result,
\begin{equation}
\begin{aligned}
    f^{A; a}_{NN}(\beta, \qv_T) =& \frac{ g_A^2I_{N N}^{A; a}}{16\pi^3f_\pi^2 }\bigg[\frac{M^2(1-\beta)}{\qv_T^2+M^2(1-\beta)^2+m_\pi^2\beta}+\frac12\log\bigg(\frac{\qv_T^2+m_\pi^2}{\mu_\chi^2}\bigg)\delta(1-\beta)\bigg].
\end{aligned}
\label{eq: final fAN}
\end{equation}
Again, in Eq. (\ref{eq: DpiDN int}), we have chosen the pole at $k^- = (\kv_T^2 +m_\pi^2 -i\epsilon)/(2k^+)$, which we interpret as setting the pion on shell, hence why Fig. (\ref{fig: A cross}) has a cut through the pion propagator.


When the pion stays on only one leg, as in Fig. (\ref{fig: A loop}), it generates a loop diagram that does not have any $\qv_T$ dependence. Like the asymmetric tadpole diagrams in the nucleon operator, this contributes only at the endpoint, $\beta \to 1$. The ``asymmetric" axial hadronic distribution function is,
\begin{equation}
\begin{aligned}
    f^{A; a, {\rm as}}_{NN}&(\beta, \qv_T) = \frac{ i g_A^2 I_{N N}^{A; a, {\rm as}}}{8f_\pi^2}\int \frac{d^4k}{(2\pi)^4}\overline{u}(P)\bigg[\slashed{n}\gamma_5 \frac{i}{k^2-m_\pi^2+i\epsilon}\frac{i(\slashed{p}-\slashed{k}+M)}{(p-k)^2-M^2+i\epsilon} \slashed{k} \gamma_5\\ 
    &- \gamma_5 \slashed{k} \frac{i}{k^2-m_\pi^2+i\epsilon}\frac{i(\slashed{p}-\slashed{k}+M)}{(p-k)^2-M^2+i\epsilon} \slashed{n}\gamma_5\bigg]u(P)  \delta((1-\beta)P^+)\delta^{(2)}(\qv_T).
\end{aligned}
\label{eq: AS axial}
\end{equation}
Notice the overall sign is different between Eq. (\ref{eq: axial TMD hadronic distribution functions}) and Eq. (\ref{eq: AS axial}). The gamma structures and propagators are the same as those for Eq. (\ref{eq: axial TMD hadronic distribution functions}), so we can use the same steps to evaluate the trace and simplify. However, now there are no constraints on the $k^+$ or ${\bf k}_T$ components, meaning we need to use dimensional regularization to evaluate the $d^4k$ integral. Doing so yields the result, 
\begin{equation}
\begin{aligned}
    f^{A; a, {\rm as}}_{NN}(\beta, \qv_T) =& \frac{-g_A^2I_{N N}^{A; a, {\rm as}}}{(4\pi)^2 f_\pi^2}\bigg[\frac{m_\pi^2}{2}\bigg(1- \log\bigg(\frac{m_\pi^2}{\mu_\chi^2}\bigg)\bigg) \\
    &- M^2\int_0^1 dx x\log\bigg(\frac{\Delta}{\mu_\chi^2}\bigg)\bigg]\delta(1-\beta)\delta^{(2)}(\qv_T),
\end{aligned}
\end{equation}
where $\Delta = (1-x)m_\pi^2 +x^2 M^2$.

\subsection{Pion-Nucleon hadronic distribution functions}
\begin{figure}[t]
\begin{subfigure}{0.35\linewidth}
\includegraphics[width = .56\linewidth]{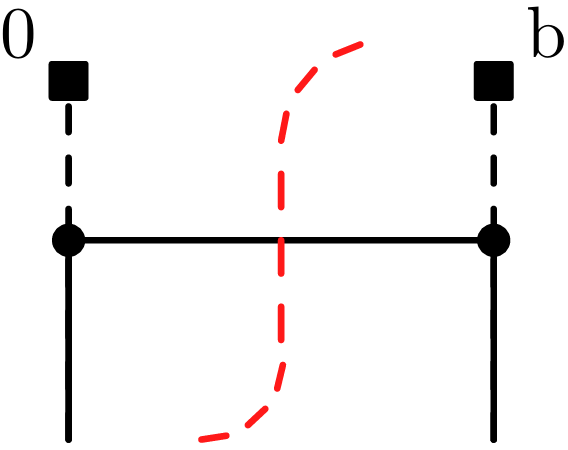}
\caption{}
\label{fig: pi axial}
\end{subfigure}
\begin{subfigure}{0.35\linewidth}
\includegraphics[width = .5\linewidth]{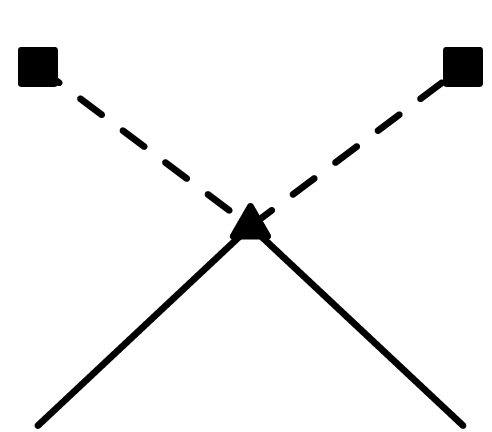}
\caption{}
\end{subfigure}
    \caption{Loop corrections to pion operators with cut. Squares represent an operator insertion, circles are the axial interaction, and the triangle is the $\pi \pi NN $ interaction.}
    \label{fig: PO diagrams}
\end{figure}

Lastly, at the order we are working, the pion operator can also contribute to the nucleon TMD PDFs. In this case, there is only one term in the operator, Eq. (\ref{eq: Opi}), but there are two possible interactions, the axial interaction and the $\pi\pi NN$ interaction, known as the Weinberg-Tomozawa coupling. As a result, ${\cal O}_\pi$ has two diagrams. These are shown in Fig. \ref{fig: PO diagrams}.

The diagram in Fig. \ref{fig: pi axial} comes from two insertions of the axial interaction. Its hadronic distribution function is given by
\begin{equation}
\begin{aligned}
    f^{a}_{\pi N}(y, \qv_T) =& \frac{ g_A^2I_{\pi N}^{a}}{8f_\pi^2}\int \frac{d^4k}{(2\pi)^4}\overline{u}(P)\slashed{k}\gamma_5 \frac{i(\slashed{P}-\slashed{k}+M)}{(P-k)^2-M^2+i\epsilon} k^+ \bigg(\frac{i}{k^2-m_\pi^2+i\epsilon}\bigg)^2\gamma_5 \slashed{k} u(P)\\
    &\times\delta(k^+ -yP^+)\delta^{(2)}(\qv_T -\kv_T).
\end{aligned}
\end{equation}
where $y = 1 - \beta$ is the momentum fraction of the pion in the nucleon and $I_{\pi N}^{a}$ is the isospin factor. Averaging over spin and taking the trace gives
\begin{equation}
\begin{aligned}
    f^{a}_{\pi N}(y, \qv_T) =& \frac{-i g_A^2I_{\pi N}^{a}}{16  f_\pi^2}\int \frac{d^4k}{(2\pi)^4}\bigg[\frac{8k^+ M^2m_\pi^2}{[(P-k)^2-M^2+i\epsilon](k^2-m_\pi^2+i\epsilon)^2}\\
    &+ \frac{8k^+ M^2}{[(P-k)^2-M^2+i\epsilon](k^2-m_\pi^2+i\epsilon)}+\frac{4k^+(P\cdot k)}{(k^2 -m_\pi^2+i\epsilon)^2}\bigg]\\
    &\times\delta(k^+ -yP^+)\delta^{(2)}(\qv_T -\kv_T).
\end{aligned}
\label{eq: traced pion}
\end{equation}
The basis integrals needed are given by Eq. (\ref{eq: DpiDN int}) and 
\begin{subequations}
\begin{equation}
\begin{aligned}
   \int dk^-  &\frac{1}{(k^2-m_\pi^2+i\epsilon)^2[(P-k)^2 - M^2+i\epsilon]} = \frac{\pi i(k^+-P^+)}{(P^+)^2[{\bf k}_T^2 +m_\pi^2(1-k^+/P^+)+M^2(k^+/P^+)^2]^2},
\label{eq: Dpi2DN int}
\end{aligned}
\end{equation}
\begin{equation}
    \int dk^- k^-\bigg(\frac{1}{k^2-m_\pi^2}\bigg)^2 = \frac{1}{2} i\pi \bigg[\frac{1}{\epsilon} - \log\bigg(\frac{{\bf k}_T^2 + m_\pi^2}{\mu_\chi^2}\bigg)\bigg]\frac{\partial}{\partial k^+}\delta(k^+),
\label{eq: Dpi2 kpkm}
\end{equation}
\begin{equation}
    \int dk^- \bigg(\frac{k^+}{k^2-m_\pi^2}\bigg)^2 = 0,
\label{eq: Dpi2 kpkp}
\end{equation}
\end{subequations}
We evaluate the first integral in a similar manner to Eq. (\ref{eq: DN Dpi2 int}). Like for the diagrams in Fig. \ref{fig: N LO} and \ref{fig: A cross}, the poles are for Fig. \ref{fig: pi axial} are two different halves of the complex plane. However, in Eq. (\ref{eq: Dpi2DN int}) the only simple pole comes from the nucleon propagator. Thus we pick $k^- = (\kv_T^2 + M^2 k^+/P^+ -i\epsilon)/(2P^+ - 2k^+)$, which sets the nucleon on-shell and can be interpreted as putting a cut through the nucleon propagator in Fig. \ref{fig: pi axial}. Equations (\ref{eq: Dpi2 kpkm}) and (\ref{eq: Dpi2 kpkp}) are evaluated by taking derivatives with respect to $k^+$ on (\ref{eq: Dpi km}) which is why we see the appearance of similar divergences. 

Plugging in the basis integrals to Eq. (\ref{eq: traced pion}), $f^{a}_{\pi N}(y, \qv_T)$ evaluates to,
\begin{equation}
\begin{aligned}
    f^{a}_{\pi N}(y, \qv_T) =& \frac{g_A^2I_{\pi N}^{a}}{32\pi^3f_\pi^2}\bigg(\frac{M^2 y(\qv_T^2+M^2y^2)}{[\qv_T^2+M^2y^2+m_\pi^2(1-y)]^2} +\frac{1}{4} \log\bigg(\frac{{\bf q}_T^2 + m_\pi^2}{\mu_\chi^2}\bigg)\delta(y) \bigg).
\end{aligned}
\label{eq: final fpiN}
\end{equation}

Lastly, the second diagram in Fig. \ref{fig: PO diagrams} has two pion propagators with a $\pi \pi NN$ vertex. The corresponding hadronic distribution function is,
\begin{equation}
\begin{aligned}
    f^{a}_{\pi \pi N}(y, \qv_T) =& \frac{- iI_{\pi \pi N}^{a}}{16f_\pi^2}\int \frac{d^4k}{(2\pi)^4}\overline{u}(P) (i \slashed{k}) \frac{i}{k^2-m_\pi^2+i\epsilon} k^+ \frac{i}{k^2-m_\pi^2+i\epsilon}u(P)\\
    &\times \delta(k^+ -y P^+)\delta^{(2)}(\qv_T -\kv_T).
\end{aligned}
\end{equation}
Averaging over spins and using Eqs (\ref{eq: Dpi2 kpkm}) and Eq. (\ref{eq: Dpi2 kpkp}), we can quickly get the result for $f^{a, tad}_{\pi N}$,
\begin{equation}
\begin{aligned}
    f^{a}_{\pi \pi N}(y, \qv_T) =& \frac{- iI_{\pi \pi N}^{a}}{32(2\pi)^3 f_\pi^2}  \log\bigg(\frac{{\bf q}_T^2 + m_\pi^2}{\mu_\chi^2}\bigg)\delta(y).
\end{aligned}
\end{equation}
Notice this contribution has nontrivial $\qv_T$ dependence but only contributes at the endpoint, $y=0$, i.e., $\beta = 1$.

\section{Results}
\label{sec: results}

Before studying the TMD hadronic distribution functions numerically, we need to first determine the various isospin factors appearing in the calculations. These isopin factors depend on the quark TMD PDF being calculated. For an up quark distribution, replace $\Gamma^a \to \frac12(1 + \tau^3)$. For the down distribution, take $\Gamma^a \to \frac12(1 - \tau^3)$. For $u-d$, replace $\Gamma^a \to \tau^3$. For the distribution in a proton, the isospin factors for $\Gamma^a$ = $1$ and  $\Gamma^a = \tau^3$ are given in Table \ref{tab: Isospin factors}.

\begin{table*}
\begin{tabular}{|c|| c | c | c | c | c | c | c |}
\hline
~$I_F$~ & ~~$I^a_{Np}$~~  & ~~$I^{a, {\rm s}}_{Np}$~  & ~~$I^{a, {\rm as}}_{Np}$~~  & ~$I^{A; a}_{Np}$ & ~~$I^{A; a, {\rm s}}_{Np}$~~ & ~~$I^a_{\pi p}$~~ & ~~$I^a_{\pi \pi p}$~~ \\
\hline
~~~$\Gamma^a$ = $1$~~~ & 3& 3& 3& 3 &3 & 6 & 0\\

~~~$\Gamma^a = \tau^3$~~~ & -1  & -1 & 3 & -1 & 3 & 4 & -4i\\
\hline

\end{tabular}
\caption{Isopin factors for diagrams with a proton external state.}
\label{tab: Isospin factors}
\end{table*} 

As a case study, we compute the isovector $u-d$ distribution in the proton. This corresponds to choosing $\Gamma^a = \tau^3$ and $\ket{N} = \ket{p}$ in Eq. (\ref{eq: nonsinglet TMD}), so we use the second row of Table \ref{tab: Isospin factors}. Following \cite{Burkardt:2012hk} we label these isovector distributions with {\it iv}. To study the hadronic distribution functions numerically, we choose the physical values for the constants: $g_A = 1.26$, $f_\pi = 0.093$ GeV, $m_\pi = 0.139$ GeV, and $M = 0.939$ GeV. 

\begin{figure}[t]
    \centering
    \includegraphics[width = .9\linewidth]{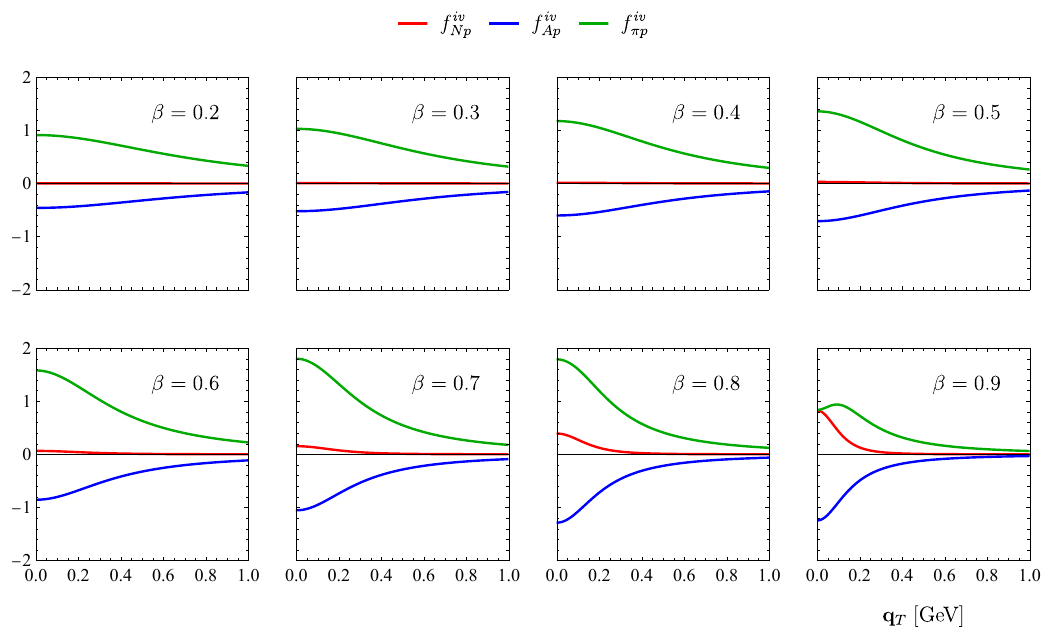}
    \caption{TMD hadronic distribution functions plotted as a function of the transverse momentum at various values of the intermediate nucleon's momentum fraction, $\beta$.}
    \label{fig: hadronic distribution function plot}
\end{figure}

We plot the hadronic distribution functions as functions of $\qv_T$ from 0 to 1 GeV at multiple values of the nucleon momentum fraction, $\beta$. These ranges are chosen to stay within the TMD regime and to stay away from the endpoint region, $\beta = 1$, to avoid the delta function contributions. Thus, the only terms that have nontrivial $\qv_T$ dependence are $f^{iv}_{Np}, f^{A; iv}_{Np},$ and $f^{iv}_{\pi p}$. The results are shown in Fig. \ref{fig: hadronic distribution function plot}.

The first observation is that $ f^{A; iv}_{Np},$ and $f^{iv}_{\pi p}$ are substantially larger in magnitude than $f^{iv}_{Np}$ for most values of $\beta$, especially for the smaller ranges. This makes intuitive sense because a small $\beta$ implies that the majority of the nucleon's longitudinal momentum is carried by the intermediate pion - hence we expect some suppression on the nucleon-nucleon hadronic distribution functions. It is not until $\beta$ becomes large that we see $f^{iv}_{Np}$ become relevant. Physically, this finding indicates that intermediate nucleon states should not substantially contribute to the long distance TMD physics of a proton, except for at high values of the nucleon momentum fraction.

It is interesting to point out that $f^{iv}_{\pi p}\approx -2 f^{iv}_{Ap}$ when $\beta$ is between 0.2 and 0.6. This is not a coincidence - the sum rule in App. \ref{app: sum rule} requires $4f^{iv}_{Ap}+2f^{iv}_{\pi p} +4f^{iv}_{Np} =0$. Since $4f^{iv}_{Np}$ is small for 0.2 $\le \beta \le 0.5$, we'd expect $-2f^{iv}_{Ap} \approx f^{iv}_{\pi p}$. We have verified that Eqs. (\ref{eq: final fNN}), (\ref{eq: final fAN}), and (\ref{eq: final fpiN}) satisfy this condition for all $\qv_T \neq 0$ and $\beta \neq 1$.

To actually make a prediction for the TMD PDFs, one needs detailed information about the valence PDFs in the pion and the polarized valence PDFs in the nucleon. A numerical implementation of the convolution between the TMD hadronic distribution functions and the valence PDFs is beyond the scope of this work. We therefore leave a  full application of the convolution in Eq. (\ref{eq: TMD matching momentum}) for future studies.

\section{Conclusion}
\label{sec: conclusion}
In this work we have developed a systematic effective field theory approach for matching TMDs onto $\chi$PT. Our main result is Eq. (\ref{eq: TMD matching}) which expresses the TMD PDF as a convolution of the valence TMD PDF and the TMD hadronic distribution functions. In Eqs. (\ref{eq: TMD split func}) and (\ref{eq: TMD chiral operators}) the TMD hadronic distribution functions are defined as Fourier transforms of nonlocal operators in the chiral theory. Analogous formulae for collinear PDFs can be derived by matching $\chi$PT onto local twist two operators. We do not derive Eqs. (\ref{eq: TMD matching})-(\ref{eq: TMD chiral operators}) but give arguments for their correctness. The nonlocal chiral operators reduce to the appropriate Noether currents in the limit of zero separation, and the nonlocal operators are the unique point splitting of the Noether currents that preserves chiral invariance. When transformed to momentum space, the resulting momementum space convolution obeys an intuitive momentum conservation relation. We explore the various limits of the convolution and speculate that the OPE can be applied to the valence TMD, allowing it to be written is in terms of a matching coefficient and the collinear PDF. Finally, we calculate the TMD hadronic distribution functions defined by Eqs. (\ref{eq: TMD split func}) to one loop in chiral perturbation theory. Our results satisfy a nontrivial sum rule. Numerical results for the TMD hadronic distribution functions as a function of $\qv_T$ for various values of $\beta$, the collinear momentum fraction of the nucleon, are also presented. In future work, we will use this formalism to study the TMD PDFs in the proton and will compare with current experimental extractions.


Some obvious future directions include incorporating the $\Delta$ resonance and extending the calculation to SU(3) chiral effective theory to include strangeness. Additionally, it will be possible to study polarized TMD distributions and the TMDs of other hadrons, like pions, in this framework. Systematic corrections can be made using this formalism as well. Incorporating higher order calculations in the chiral expansion ($p_\pi/\Lambda_\chi, m_\pi/\Lambda_\chi$), TMD expansion ($\Lambda_{QCD}^2/\pv_T^2$), and in perturbative QCD ($\alpha_s$), should all improve the accuracy of the predictions and would be worthwhile efforts. It also seems necessary to explore rapidity divergences further in the formalism. We also note that, while we expect their effects to be small, higher order chiral operators that renormalize the operators in this paper should also be studied for completeness. This paper serves to initiate the application of $\chi$PT to the TMD parton distributions functions. It is clear that there are many possible ways to expand on what has been done and all of these directions should be explored in the future.
%
%

\acknowledgements We thank Jyotirmoy Roy, Berndt Mueller, Wally Melnitchouk, Chueng Ji, Johannes Michel, Alexey Vladimirov, Michael Luke, and Iain Stewart for helpful conversations. M.~C., and T.~M. are supported by the U.S. Department of Energy, Office of Science, Office of Nuclear Physics under grant contract Numbers  DE-FG02-05ER41367. T.~M. is also supported by  the Topical Collaboration in Nuclear Theory on Heavy-Flavor Theory (HEFTY) for QCD Matter under award no.~DE-SC0023547. M.C. is supported by the National Science Foundation Graduate Research Fellowship under Grant No.~DGE 2139754. 

\appendix

\section{Sum rule for hadronic distribution functions} 

\label{app: sum rule}

We can use the matching in Eq. (\ref{eq: TMD matching momentum}) to derive a simple relation between the TMD hadronic distribution functions. Similar relations were found to hold for the collinear hadronic distribution functions in Refs. \cite{Burkardt:2012hk, Ji:2013bca}. For clarity, we derive a relation for the $u-d$ distribution in the proton. As explained in Sec. \ref{sec: results}, this is the isovector distribution so we  label the distributions with $iv$. When we write various contributions from the operators, we find
\begin{equation}
\begin{aligned}
    f_{(u-d)/p}(\zeta, \kv_T) = & Z_2 (u-d)^{v}_p + \bigg[ (u-d)^{v}_{\pi^+} \otimes \bigg( f^{iv}_{\pi p} +  f^{iv}_{\pi \pi p}\bigg)+ (u-d)^{v}_{p} \otimes \bigg( f^{iv}_{N p} +  f^{iv, s}_{N p} +  f^{iv, as}_{N p}\bigg) \\
    &  + \frac{1}{g_A}(\Delta u-\Delta d)^{v}_{p} \otimes \bigg( f^{iv, A}_{N p} + f^{iv, A, {\rm as}}_{N p}\bigg)\bigg],
\label{eq: explicit TMD matching}
\end{aligned}
\end{equation}
where each term depends on $\zeta$ and $\kv_T$ and the convolutions are in the momentum fraction and in transverse momentum, as in Eq. (\ref{eq: TMD matching momentum}). In Eq. (\ref{eq: explicit TMD matching}), $Z_2$ is the wave function renormalization. It can be determined at one loop via \cite{Ji:2013bca},

\begin{equation}
\begin{aligned}
    &(Z_2 -1) \approx (1-Z_2^{-1}) = \frac{1}{2 P^+} \frac{\partial \Sigma~}{~\partial P^-}\\
    &= \frac{1}{2P^+} \frac{-3g_A^2}{(2f_\pi)^2}\int \frac{d^4 k}{(2\pi)^4} \overline{u}(P) \slashed{k}\gamma_5 \frac{i(\slashed{P}-\slashed{k}+M)}{(P-k)^2 - M^2 +i\epsilon}\frac{i}{k^2-m_\pi^2+i\epsilon} \\
    &\times\slashed{n} \frac{i(\slashed{P}-\slashed{k}+M)}{(P-k)^2 - M^2 +i\epsilon} \gamma_5 \slashed{k} u(P).
\end{aligned}
\end{equation}
If you make the substitutions $k^+ \to (1-\beta)P^+$ and $\kv_T \to \qv_T$, 
\begin{equation}
\begin{aligned}
    &(Z_2 -1) \approx \frac{1}{2P^+} \frac{\partial \Sigma~}{~\partial P^-}\\
    &= \frac{1}{2} \frac{-3g_A^2}{(2f_\pi)^2}\int d\beta d^2\qv_T 
 \int \frac{d k^-}{(2\pi)^4} \overline{u}(P) \slashed{k}\gamma_5 \frac{i(\slashed{P}-\slashed{k}+M)}{(P-k)^2 - M^2 +i\epsilon}\frac{i}{k^2-m_\pi^2+i\epsilon} \\
    &\times\slashed{n} \frac{i(\slashed{P}-\slashed{k}+M)}{(P-k)^2 - M^2 +i\epsilon} \gamma_5 \slashed{k} u(P)\bigg |_{k^+ = (1-\beta)P^+, \kv_T = \qv_T},
\end{aligned}
\end{equation}
then by comparison with Eq. (\ref{eq: NN feyn}) we see that 
\begin{equation}
    (Z_2 -1) = 3 \int  d\beta d^2\qv_T f^{iv}_{Np}(\beta, \qv_T).
\end{equation}
Plugging this in to Eq. (\ref{eq: explicit TMD matching}) and integrating over the momentum fraction and the transverse momentum gives (at leading order in QCD perturbation theory),
\begin{equation}
\begin{aligned}
    0 = \int d\beta d^2\qv_T \bigg[ 2 \bigg( f^{iv}_{\pi p} +  f^{iv}_{\pi \pi p}\bigg)+  \bigg( 4f^{iv}_{N p} +  f^{iv, s}_{N p} +  f^{iv, as}_{N p}\bigg) +  \bigg( f^{iv, A}_{N p} + f^{iv, A, {\rm as}}_{N p}\bigg)\bigg],
\label{eq: TMD sum rule}
\end{aligned}
\end{equation}
where we have used $\int d\zeta d^2\kv_T f_{(u-d)/p}(\zeta, \kv_T)=1, \int d\alpha d^2\pv_T \big[u^v_p(\alpha, \kv_T) - d^v_p(\alpha, \kv_T)\big] = 1, \int d\alpha d^2\pv_T \big[ u^v_{\pi^+}(\alpha, \kv_T) - d^v_{\pi^+}(\alpha, \kv_T)\big] = 2 $ and $\int d\alpha d^2\pv_T \big[ \Delta u^v_p(\alpha, \kv_T) - \Delta d^v_p(\alpha, \kv_T)\big] = g_A$. This equation provides an explanation for the relation between diagrams observed numerically in Sec. \ref{sec: results}. Since only $f^{iv}_{\pi p}, f^{iv}_{N p}, f^{iv, A}_{N p}$, and $f^{iv, A, l}_{N p}$ contain nucleon propagators we expect that their contributions away from $\beta \to 1$ need to cancel out exactly to preserve Eq. (\ref{eq: TMD sum rule}). Noting that 
\begin{equation}
    \int d\beta d^2\qv_T f^{iv, A}_{N p} =  3 \int d\beta d^2\qv_T f^{iv, A, {\rm as}}_{N p},
\end{equation}
we find the expected relationship between the plotted diagrams is given by
\begin{equation}
\begin{aligned}
    0 = \int d\beta d^2\qv_T \bigg[ 2 f^{iv}_{\pi p} + 4f^{iv}_{N p}  + 4 f^{iv, A}_{N p} \bigg],
\label{eq: TMD sum rule}
\end{aligned}
\end{equation}
which is exactly what we observe in Sec. \ref{sec: results}. 
\bibliography{main}

\end{document}